\documentclass[12pt,aps]{revtex4}
\usepackage{epsfig,amsmath,amssymb}

\newcommand{\fr}[1]{
             \frac{#1}}
\newcommand{\bea}{\begin{eqnarray}}
\newcommand{\eea}{\end{eqnarray}}
\newcommand{\beq}{\begin{equation}}
\newcommand{\eeq}{\end{equation}}

\newcommand{\ket}{{\rangle }}

\newcommand{\bra}{{\langle }}
\newcommand{\gc}{\bra\fr{\alpha_s}{\pi}G^2\ket}

\newcommand{\qc}{\bra\,\overline{q}q\,\ket}

\begin{document}

\title{Non-leptonic decays in an extended \\ chiral quark model\footnote{Presented at QCD@work, Lecce, Italy, june 2012,Published in AIP Conf.Proc. 1492 (2012) 88-94}}

\author{J.O. Eeg}

\affiliation{Department of Physics, University of Oslo, P.O. Box 1048
Blindern, N-0316 Oslo, Norway}

\vspace{0.5cm}


\begin{abstract}

We consider the 
 color suppressed (nonfactorizable) amplitude for the decay 
 mode $\overline{B_{d}^0} \rightarrow \, \pi^0 \pi^{0} $. 
We treat the  $b$- quark  in the heavy quark limit  and the
 energetic light ($u,d,s$) quarks  within a variant of
 Large Energy Effective Theory   combined  with an extension
 of chiral quark models.

Our  calculated 
 amplitude for 
$\overline{B_{d}^0} \rightarrow \, \pi^0 \pi^{0} $ is suppressed
 by a factor of order
$\Lambda_{QCD}/m_b$ with respect to the factorized amplitude, as it should
according to QCD-factorization.
 Further, for 
 reasonable values of the (model dependent) gluon condensate and the 
constituent quark mass,  the calculated  nonfactorizable amplitude for 
  $\overline{B_{d}^0} \rightarrow \, \pi^0 \pi^{0} $ can easily accomodate 
the  experimental value. 
Unfortunately, the color suppressed  amplitude is very sensitive to  
the  values of these model dependent parameters. Therefore fine-tuning is
 necessary in order to obtain an amplitude compatible with the experimental
 result for $\overline{B_{d}^0} \rightarrow \, \pi^0 \pi^{0} $.

\end{abstract}

\maketitle

\maketitle

\vspace{1cm}

Keywords: $B$-decays, factorization, gluon condensate

 PACS: 13.20.He, 12.39.Fe, 12.39.Hg, 12.39.St



\section{Introduction}

The  decay modes of the type 
$B \rightarrow \pi \, \pi $ are dynamically different. 
For the case  $\overline{B_{d}^0} \rightarrow \, \pi^+ \pi^- $
there is a substantial factorized contribution which dominates.
In contrast,  the decay mode 
  $\overline{B_{d}^0} \rightarrow \, \pi^0 \pi^{0} $ 
has a small  factorized contribution, being proportional to a  small Wilson
  coefficient combination. However,
for  the decay mode
  $\overline{B_{d}^0} \rightarrow \, \pi^0 \pi^{0} $
there is a  sizeable  nonfactorizable (color suppressed) contribution due to 
soft (long distance) interactions,
which
dominate the amplitude.

In spite of tremendous effort
within QCD factorization \cite{Beneke:1999br},  soft collinear effective theory (SCET),
\cite{Bauer:2000ew,Bauer:2000yr}, the socalled pQCD model \cite{Li:2003az}, and 
 QCD sum rules, the obtained theoretical 
amplitude for $\overline{B_d^0} \rightarrow \pi^0 \pi^0 $
is still a factor $\sim$ 2 off \cite{Khodjamirian:2005wn}.

Chiral quark models bridge between 
the  $quark$ and $meson$ picture and has shown to be  suitable
 for calculating color suppressed decays \cite{Eeg:1992bi,Eeg:1993br,Bertolini:1994qk,Bertolini:1997ir,Bertolini:1997nf,Bergan:1996sb,Eeg:2001un,Eeg:2003yq,Eeg:2005au,Eeg:2001tg,Eeg:2005bq,Eeg:2004ik,MacdonaldSorensen:2006ds,Eeg:2003pk}.
In this talk I  report on a  calculation \cite{Eeg:2010rk} of  
 $\overline{B_d^0} \rightarrow \pi^0 \pi^0 $ within a 
extension of chiral quark model \cite{Leganger:2010wu}.

\section{Effective theories}


The effective non-leptonic quark level Lagrangian at
  a scale $\mu$ has the form: 
 \beq
 {\mathcal L}_{W}=  \sum_i  C_i(\mu) \; \hat{Q}_i (\mu)
\; ,
 \nonumber
\eeq
where
  $C_i$  are Wilson coefficients containing loop effects from 
scales above $\mu$.
Typically, the operators  $\hat{Q}_{i}$'s  are products of two left-handed 
quark currents.
Genericlly, for non-leptonic processes 
with two numerically relevant operators 
$\hat{Q}_{X,Y}$ one obtains
 \begin{eqnarray}
\langle M_1 \, M_2| {\mathcal L}_{W} | M \rangle \, = \, 
   \left(C_X \, + \frac{C_Y}{N_c} \right) \, \langle M_1| \, j_L^\alpha(1) |
   0 \rangle \, \langle M_2 | j_\alpha^L(2) | M \rangle \, 
+ \,  C_Y \, \langle M_1 \, M_2| \, \hat{Q}_Y^{color} \, | M \rangle \; .
\end{eqnarray}
Here, for  ``flavor mismatch'', as shown in the right-hand  diagram of fig.2,
 one has  used a 
  Fierz transformation. The operator $\hat{Q}_Y^{color}$ is a product of 
two colored currents of the type
$j^a_\alpha = \overline{q_1} \gamma_\alpha L t^a q_2$, where $q_{1,2}$ are 
quark fields,  $t^a$ a color matrix and $L$ the left-handed projector
 in Dirac space.

In some cases the coefficient combination 
 $ \left(C_X \, + \frac{C_Y}{N_c} \right)$ is
 close to zero. Then matrix elements of $\hat{Q}_Y^{color}$ might dominate.
It is important to note that the  matrix elements of colored operators
 might be calculated within chiral quark models.


\begin{figure}[t]
\begin{center}
   \epsfig{file=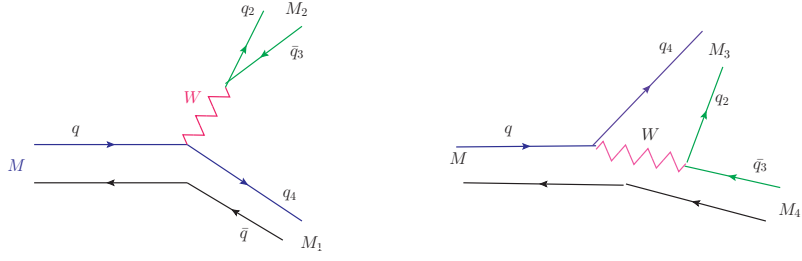,width=11cm}
\label{fig:NonLeptFig}
\caption{Factorizable (``color allowed'') and color-suppressed (right)
diagrams}
\end{center}
\end{figure}

 Heavy  ($b$ eventually also $c$-) quarks might be  treated within
 Heavy Quark Effective Theory ($HQET$) \cite{Neubert:1993mb,Manohar:2000dt}.
In this theory one projects out the movement of a heavy quark :
 $p_Q = m_Q \, v \, + \, k \;$, where the heavy quark  velocity $v$ satisfies  
 $v^2=1$, and $m_Q$ is the heavy quark mass.
 One obtains an effective Lagrangian for the reduced heavy quark
 field $Q_v$ and the corresponding propagator $S(p_Q)$:
\begin{eqnarray}
{\cal L}_{HQET}= 
 \overline{Q_{v}} i v \cdot D Q_{v}
 + {\cal O}(m_Q^{- 1}) \; \;  ;  \; \; 
 S(p_Q) 
\; = \; \frac{(1 +  \gamma \cdot v)}{2 \, k \cdot v} \; \, .
\end{eqnarray}
`

For energetic light ($u,d,s$ ) quarks in the final state 
one might use
 {\em Large Energy Effective Theory} \cite{Dugan:1990de,Charles:1998dr} 
which was
 later developed into SCET \cite{Bauer:2000ew,Bauer:2000yr}.
In this case one 
projects out movement of light energetic quark with momentum
$p^\mu_q = E \, n^\mu + k^\mu $, where \cite{Leganger:2010wu}
 \begin{eqnarray}
n ( \, {\mbox or \; \tilde{n}}) \, = \, (1,0,0, \pm \eta)
\quad ; \; \, 
\eta = \sqrt{1-\delta^2} \, , \;
 \;  n^2 = \tilde{n}^2 = \delta^2 \,
, \,
v\cdot n = v\cdot \tilde{n} = 1.
\end{eqnarray}
Here $\delta \sim \Lambda_{QCD}/m_Q \ll 1$.
In the original version \cite{Charles:1998dr}, 
 $\delta = 0$, but it was found later\cite{Leganger:2010wu}
that it was necessary to have  $\delta \neq 0$ in order
 to combine LEET with chiral quark models.
One obtains an effective Lagrangian for reduced light energetic
 quark field $q_n$: 
\begin{eqnarray}
 {\cal L}_{LEET\delta} \, = \,
 \bar{q}_n \left(\frac{1}{2}(\gamma \cdot  \tilde{n} + \delta) \right)
(i n  \cdot D) \, q_n
   + \mathcal{O}(E^{-1}) \; \; ;  \; \;
 S(p_q) \; = \;  
  \frac{\gamma \cdot n}{2 n \cdot k} \; \, .
\end{eqnarray}
In the formal limits $M_H \rightarrow \infty$ and 
$E\rightarrow\infty$,
$\langle P \, | \, V^\mu \, | \, H \rangle$ of the form \cite{Charles:1998dr}:
\begin{equation}
\langle P| V^\mu |H \rangle = 2E\left[\zeta^{(v)}(M_H,E) \, n^\mu 
                      + \zeta^{(v)}_1(M_H,E) \, v^\mu  \right] \; , 
\label{HECurrent}
\nonumber
\end{equation}
where the form factors scale as:
\begin{equation}
\zeta^{(v)}   = C \frac{\sqrt{M_H}}{E^2} \quad , 
\; \, C \sim (\Lambda_{QCD})^{3/2} \qquad , \quad 
\frac{\zeta^{(v)}_1}{\zeta^{(v)}} \sim \delta \; \sim \; \frac{1}{E} \; \; . 
\label{eq:charlesEq}
\end{equation}
The behavior  of the form factors are constistent  with the energetic quark
 having  $x$  close to one, where $x$ is the  quark 
 momentum fraction  of the outgoing pion.

 Within the mesonic picture one has low energy effective theories, 
mainly chiral 
perturbation theories, both for the pure light sector and the
 heavy-light sector , with effective cut-offs
 $\Lambda_{\chi} \, \sim \,  4 \pi \, f_\pi \, \sim \, 1$  GeV. 
Such effective theories contain the light meson fields 
($\pi,K,\eta_8$) in a 3 by 3  matrix $\Pi$. 
The chiral Lagrangians will contain the fields \cite{Casalbuoni:1996pg} 
\begin{equation}
\xi \equiv \exp \left( \frac{i}{2 f_\pi} \, \Pi  \right) \quad ; \quad 
H^{(\pm)} =  P_{\pm}(v) \,  (P_{\mu}^{(\pm)} \gamma^\mu -     
i P_{5}^{(\pm)} \gamma_5) \; \, ,
\end{equation}
where $H_v$ is the combined field consisting of one light and one heavy 
($b$- eventually also $c$-quark) with spin-parity $0^-$ and $1^-$ .

\section{Chiral quark models}
 
\begin{figure}[t]
\begin{center}
   \epsfig{file=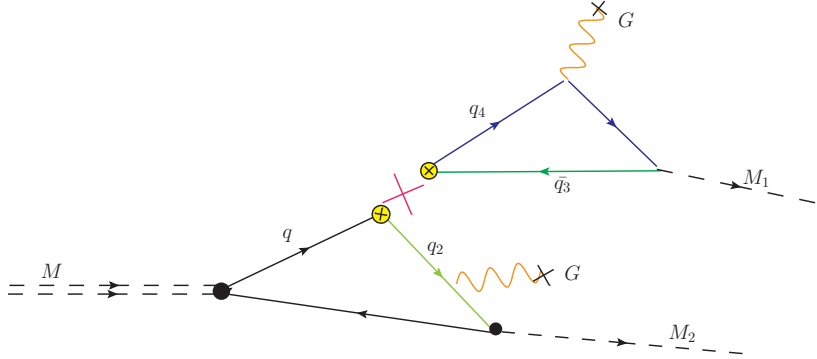,width=11cm}
\caption{Bosoniztion of colored operator $\hat{Q}^{color} \; 
 \rightarrow \; (1/N_c )\; Tr(...) \times Tr(...)$}
\label{fig:colsuppr}
\end{center}
\end{figure}

Chiral quark models are the bridge between the quark and the meson pictures. 
Loop momenta in $\chi$QM's should not exceed the chiral symmetry breaking
 scale  $ \Lambda_\chi$. In the pure   light ($q=u,d,s$)
 sector the Lagrangian might be written 
\cite{Manohar:1983md,Espriu:1989ff,Pich:1990mw,Andrianov:1991te,Bijnens:1993ap}:
\begin{equation}
{\cal L}_{\chi QM} =  
\overline{\chi} \left[\gamma^\mu (i D_\mu + {\cal V}_{\mu}  +  
\gamma_5  {\cal A}_{\mu}) -  m \right]\chi \, + \, {\cal O}(m_q)
 \nonumber
\end{equation}
where $m$ is the   {\em constituent} light quark mass, due to chiral
 symmetry breaking. 
This implies  meson-quark couplings modelling confinement!
The quark fields $\chi$ are flavor rotated versions of the ordinary
 left and righthanded quark fields $q_{L,R}$, namely:
$ \chi_L =  \xi  q_L \; \,  , \; \;  \chi_R =  \xi^\dagger q_R \; $. 
A color suppressed  suppressed decay $M \rightarrow M_1 \, M_2$
 will within chiral quark models 
generically look like in Fig \ref{fig:colsuppr} .


An important ingredient in our treatment of color suppressed amplitudes is 
the emision of soft gluons making (model dependent) gluon
 condensates \cite{Novikov:1983gd}:
\begin{equation}
g_s^2 G_{\mu \nu}^a G_{\alpha \beta}^a  \; \rightarrow 4 \pi^2
 \gc \frac{1}{12} (g_{\mu \alpha} g_{\nu \beta} -  
g_{\mu \beta} g_{\nu \alpha} ) \; \, . 
\end{equation}
Within chiral quark models the 
logarithmic-  and quadratic divergent
 integrals  are identified with the physical $f_\pi$ and the
 quark condensate, respectively \cite{Manohar:1983md,Espriu:1989ff,Pich:1990mw,Andrianov:1991te,Bijnens:1993ap,Hiorth:2002pp}.

 For the heavy light case  ($HL\chi QM$), including heavy quarks
our description includes ${\cal L}_{HQET}$, ${\cal L}_{\chi QM}$
and an additional the meson-quark interaction \cite{Hiorth:2002pp,
Bardeen:1993ae,Deandrea:1998uz,Polosa:2000ym,Ebert:1994tv}:
\begin{equation}
{\cal L}_{Int}  =   
 -   G_H \,  \overline{Q_{v}} \, H_{v} \, \chi  \; + \; h.c  \; \, .
\end{equation}
Integrating out quarks (by loop diagrams) 
 should give the known HL$\chi$PT terms with definite constants in front.
The physical and model depedent paramameters $f_\pi, \qc, f_H, g_A$ are 
linked to 
(divergent) loop integrals as in the  pure light  $\chi$QM.
A fit  in strong sector gives \cite{Hiorth:2002pp}:  
$m \sim 220$ MeV, $\gc^{1/4} \sim  315 $ MeV, 
 $ \; {G_H}^2 = \frac{2 m}{f^2} \, \rho \; $ 
 where $\rho \sim 1$ and $\rho$ depend on $f_\pi,\gc, m , g_A$.  
An  ideal case for our method is $B -\overline{B} \mbox{-mixing}$,
where  a result very close to the lattice result
 was obtained \cite{Hiorth:2003ci}.

Recently, chiral quark models are extended \cite{Leganger:2010wu} 
to incorporate also light energetic quarks described by LEET.
The meson-quark interaction is then assumed to have the form
(- the light meson field $M$ is a 3 by 3 matrix):
\begin{eqnarray}
{\cal L}_{intq} =  G_A 
\bar{q} \, \gamma_\mu \gamma_5(\partial^\mu
  M) \, q_n \, + \, h.c \quad .
\end{eqnarray}
For  mesons  containing a reduced field $q_n$ we use a corresponding
 meson field $M_n$. 


\begin{figure}[t]
\begin{center}
   \epsfig{file=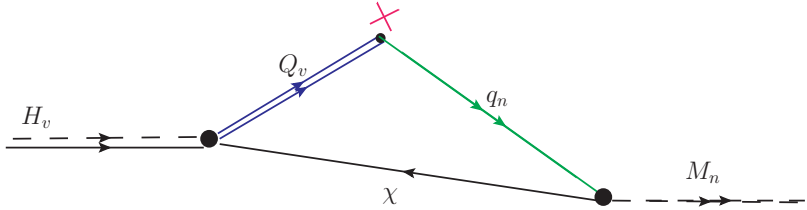,width=11cm}
\caption{ Bosonization of the heavy meson ($H_v$) to light energetic
meson ($M_n$) current (vertex in red) involving the the heavy 
reduced quark field ($Q_v$ in blue), the light energetic quark field
($q_n$ in green) and the soft  light (flavor rotated) quark field 
($\chi$ in black )}
\end{center}
\end{figure}

The coupling $G_A$ is determined by a loop diagram for $\zeta^{(v)}$
in Fig. 3. 
We find that the obtained ratio of formfactors obtained in 
 LE$\chi$QM  is fulfilling the requirements of \cite{Charles:1998dr}
in eq. (7).
The bosonized current for  $H_v \rightarrow M_n$ transition is
($H_v$ and  $M_n$ represent the heavy B- and the energetic $\pi$-meson):
\begin{equation}
J_{tot}^\mu(H_v \rightarrow M_n) \, = \;
 \left(- i \frac{G_H \, G_A}{2} \,m^2 \, F \,\right) 
   \textrm{Tr} \left\{ 
\gamma^\mu L 
H_v 
\left[\gamma \cdot n \right]
\xi^\dag M_n \right\} \; ,
\label{eq:Jmutot}
\end{equation}
where  $F= N_c/(16\pi) +...\sim 10^{-1}$.
Using 
 $\delta=m/E$, which is the chiral quark model version of 
$\Lambda_{QCD}/m_b$, we obtain a result of the type
 (recalling $M_B \simeq 2 E$ in eq. (7)):
 \begin{equation}
G_A \; \sim \; 
  \frac{1}{N_c} \, \frac{1}{E^{\frac{3}{2}}}
\label{GAExpr2}
\nonumber
\end{equation}
The coupling $G_A$ is then fixed 
from light cone sum rules:  $\zeta^{(v)} \simeq 0.3$ 
\cite{Ball:1998tj,DeFazio:2007hw}.
Using $G_A$ and $\zeta^{(v)}$ it was found  \cite{Leganger:2010wu}
that
the nonfactorizable amplitude accounted for   2/3 
of the experimental amplitude for 
$\overline{B^0}  \rightarrow   D^{0} \pi^0$.
There are also  additional meson contributions.

Our prescription for calculating non-leptonic amplitudes is then the following:
Integrate out  $W$ and heavy quarks to obtain eq. (1). Then, bosonize
 the quark operators by integrating out the  quark fields,
 and obtain an effective (chiral) Lagrangian
 at meson level. This is however often an idealized situation.
 In reality some meson
 loops cannot be calculted as chiral loops in the ordinary sense. They
might be suppressed  by  say $1/M_B$ or must be treated by other methods. 
In any case, final state interactions are in general
 present \cite{Fajfer:2005mf,Kaidalov:2006uf}.

\section{The amplitude  for 
$\overline{B^0}  \rightarrow  \pi^0 \pi^0$ in LE$\chi$QM}

Now we use the results of \cite{Leganger:2010wu} to calculate the color 
suppressed amplitude for $\overline{B^0}   \rightarrow  \pi^0 \pi^0$
 \cite{Eeg:2010rk}.


\begin{figure}[t]
\begin{center}
   \epsfig{file=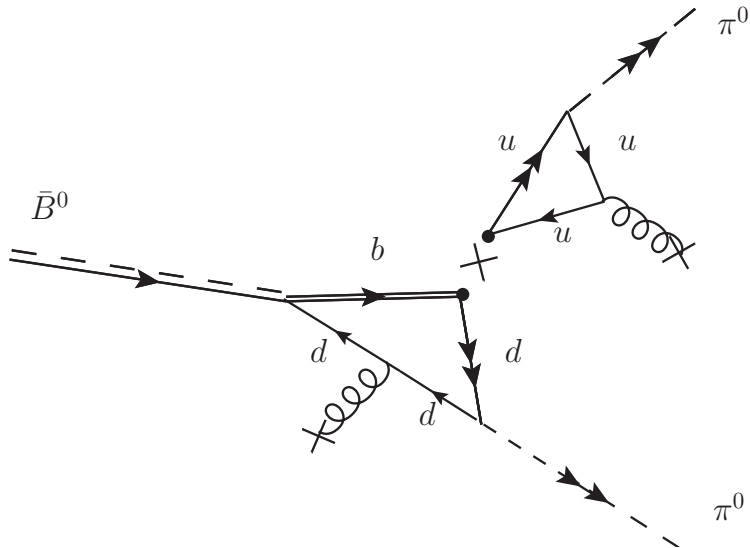,width=11cm}
\caption{Non-factorizable contribution to $B \rightarrow \pi^0 \pi^0$
 containing large energy
          light fermions and mesons. Also corresponding diagram where 
the outgoing anti-quark $\overline{u}$ is hard.}
t
\end{center}
\end{figure}

The colored $H_v \rightarrow M_n$ current (representing
 $B \rightarrow \pi$(hard))
is found to be:
\begin{eqnarray}
  J^\mu_{1G}(H_b \rightarrow M)^a \, = \,
        g_s \, G^a_{\alpha \beta} \,
          \frac{G_H \, G_A}{128 \pi}
 \epsilon^{\sigma \alpha \beta \lambda} \, n_\sigma
\mathrm{Tr} \left( \gamma^\mu L H_v
  \gamma_\lambda \,
    \xi^\dag M_n \right) \; ,
\label{eq:HtoM4}
\end{eqnarray}
Similarly the colored  current for outgoing hard $M_{\tilde{n}}$
(representing the hard pion $\pi_{\tilde{n}}$) is:
\begin{equation}
  J^\mu_{1G}(M_{\tilde{n}})^a
\, = \,   g_s G^a_{\alpha \beta} \; 
   2 \, \left(-  \frac{G_A E}{4} \right) \, Y \; 
     \tilde{n}_\sigma\epsilon^{\sigma \alpha \beta \mu} \,   
 \mathrm{Tr}\left[\lambda^X \,  M_{\tilde{n}} \right] \; \; , 
\label{eq:D0}
\end{equation}
where  $\lambda^X$ = 
appropriate  SU(3) flavor matrix, and $Y$ is a  loop factor:
\begin{equation}
    Y \,  = \, 
 \frac{f_\pi^2}{4m^2 N_c} \left( 1 \, - \,  
\frac{1}{24m^2 f_\pi^2}\langle\frac{\alpha_s}{\pi}G^2\rangle \right) \; .
\label{Q-factor}
\end{equation}

The ratio of non-factorizable to factorizable amplitudes is found to be
\cite{Eeg:2010rk} (refering to eq. (2) for the Wilson coefficient): 
\begin{eqnarray}
r \; \equiv \;
 \frac{{\cal M}(\overline{B^0_d} \rightarrow \pi^0 \pi^0)_{\mbox{Non-Fact}}}
{{\cal M}(\overline{B^0_d} \rightarrow \pi^+ \pi^-)_{\mbox{Fact}}} \; = \; 
\left(\frac{C_Y \; \sigma}{ (C_X + C_Y/N_c)} \, \right) \frac{1}{N_c} \, 
\frac{m}{E} \; \; ,
\label{ratio}
\end{eqnarray}
where $\sigma $ is a  model-dependent hadronic factor, dimension-less 
and $\sim (N_c)^0$.
Our calculations show that the ratio $r \sim 1/N_c$ 
and $r \sim m/2E \simeq \Lambda_{QCD}/m_b $ as it should according to
QCD factorization \cite{Beneke:1999br}. The ratio is plotted in Fig 5.


\begin{figure}[t]
\begin{center}
   \epsfig{file=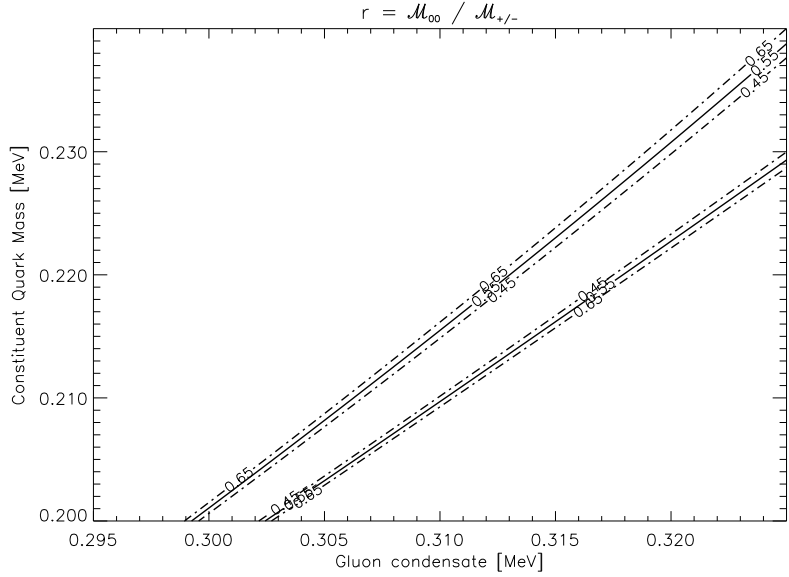,width=11cm}
\caption{Plot for  the ratio $r$
in terms of $m$ and $\gc^{1/4}$.
 For reasonable values
 of these parameters the ratio $r$ can take a wide range of values 
such that fine-tuning is required to reproduce the 
experimental value.}                   
\label{fig:Bpipi-plot}
\end{center}
\end{figure}

The experimental value of  the
$ \overline{B^0_d} \rightarrow \pi^0\pi^0$ amplitude can be  accomodated
for $m \sim$  220 Mev and  $\gc^{1/4} \sim$  315 MeV.
 - as  in previous work  \cite{Eeg:1992bi,Eeg:1993br,Bertolini:1994qk,Bertolini:1997ir,Bertolini:1997nf,Bergan:1996sb,Eeg:2001un,Eeg:2003yq,Eeg:2005au,Eeg:2001tg,Eeg:2005bq,Eeg:2004ik,MacdonaldSorensen:2006ds,Eeg:2003pk}.
But the result is  very
 sensitive to variations of  $m$ and $\gc$,
as seen by loop factor $Y$. 
In addition there are  meson loops not yet calculated.


\begin{figure}[t]
   \epsfig{file=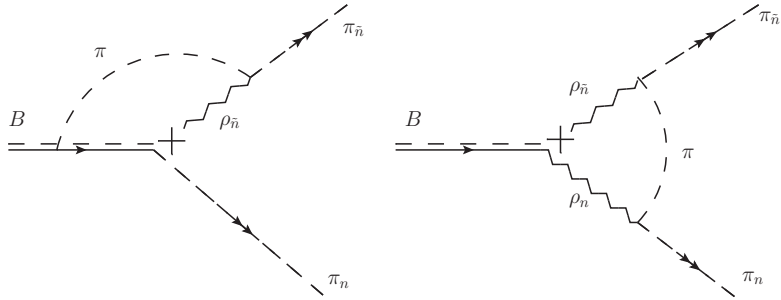,width=11cm}
\caption{Suppressed(left) and non-suppressed(right) meson loops
 for $\overline{B^0_d} \rightarrow \pi \pi$.}
\label{fig:meson-loops}
\end{figure} 

 There is an extension of the models under construction in order to
 incorporate meson loops including (energetic) vector mesons ($V_n$) in loops
and we will need 
 couplings for $V_n \rightarrow M_n + (\mbox{soft} \; \, \pi)$.
One might also
include  light non-energetic vector mesons. 

\section{ Conclusions}

We have considered the amplitude  $B^0 \rightarrow \pi^0 \overline{\pi^0}$
 within
an extended chiral quark model  LE$\chi$QM which is in accordance with
the requirements of \cite{Charles:1998dr}.

For the color suppressed $\sim 1/N_c$ part of the 
$\overline{B_d^0} \rightarrow  \pi^0 \pi^0$ amplitude , we have 
found a ratio $r \; $ 
  $\sim m/2E \simeq \Lambda_{QCD}/m_b$ suppression as we
 should according to \cite{Beneke:1999br}. Unfortunately
the  obtained amplitude is very sensitive to $m$ and $\gc$.

Extension of the model including energetic light vectors, in order to
be able to  consider
 meson loops, are under construction.

{\acknowledgments}

 I am  supported in part by the Norwegian
 research council (HEPP project)



\bibliography{lecceRef}

\end{document}